\documentclass[12pt]{article}
\usepackage[dvips]{graphicx}

\def\lromn#1{\uppercase\expandafter{\romannumeral#1}}

\begin{document}

\begin{center}
\begin{large}
\textbf{
Effect of Relic Neutrino on Neutrino Pair Emission
from Metastable Atoms
}

\end{large}
\end{center}

\vspace{2cm}
\begin{center}
\begin{large}
Toru Takahashi and
M. Yoshimura

Center of Quantum Universe and
Department of Physics, Okayama University, \\
Tsushima-naka 3-1-1, Okayama 700-8530,
Japan 
\end{large}
\end{center}

\vspace{2cm}


\begin{center}
\begin{Large}
{\bf ABSTRACT}
\end{Large}
\end{center}

A possiblity of measuring the cosmic neutrino temperature
$\sim 1.9 K$ and other important quantities such as
the chemical potential $\mu$ and the decoupling temperature $T_d$ 
is discussed, using the recently proposed process of
photon irradiated neutrino
pair emission from metastable atoms.
The Pauli blocking effect of relic neutrinos
reduces the rate by a large factor $\approx (1 + m_1/T_d)/4$
at the threshold of the lightest neutrino pair (of mass $2m_1$).
Correction of linear order in $\mu$ 
near the mass thresholds
can be used to improve the constraint on the lepton asymmetry.

\newpage
Observation of the cosmic neutrino background is a direct indication
of the hot early universe, three mintutes after the big bang.
One cannot overemphasize its importance.
It would be very exciting to explore a possibility of
measuring the relic cosmic neutrino expected to
have the Fermi-Dirac distribution of nearly zero chemical
potential and temperature $\sim (4/11)^{1/3} T_0 \approx 1.9 K$,
with $T_0 \sim 2.7 K $ the cosmic microwave background temperature
\cite{weinberg-cosmology}.

In this work we study a possibility of using
the recently proposed laser (or microwave, which is not
discussed in the present work)
irradiated neutrino pair emission from metastable atoms \cite{pair emission}
in order to indirectly detect the relic neutrino.
It is an indirect detection because the presence of relic neutrinos
is felt only by the Pauli blocking effect in this proposal.
A great merit of atomic transition is obvious;
closeness of the atomic energy level difference to neutrino masses.
The energy difference is also made close to the cosmic neutrino 
temperature;
$k_B T \sim 0.166 \,meV (T_B/ 1.9 K)$.
We examine the pair emission rate affected
by the Pauli blocking.
In the rest of this paper we take the natural unit;
$\hbar =1 \,, c= 1 \,, k_B =1$.

\begin{figure}[htbp]
  \centerline{\includegraphics[scale=0.7]{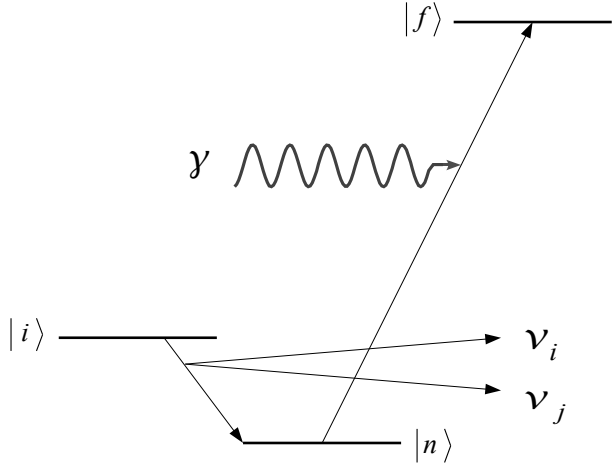}}
  \caption{Atomic level structure and laser irradiated neutrino pair emission}
  \label{Fig 1}
\end{figure}

The photon irradiated neutrino pair 
(all 6 Majorana pair channels $\nu_i \nu_j$ added) 
emission from a metastable atom proceeds as depicted in Figure 1.
The intermediate atomic state $| n \rangle $ is taken close 
in the energy to the initial metastable state $| i \rangle$ $-$ the mass
of the neutrino pair (neutrino eigenmasses arranged by $m_1 < m_2 < m_3$),
thus $E_i - E_n \approx m_i + m_j$, along with the
laser tuning condition to the final excited state $| f \rangle$, 
$\omega \approx E_{f} - E_n$. 
Thus, all 6 thresholds corresponding to
$\nu_i \nu_j$ pair emission appear at laser energies,
$\omega = E_{f} - E_i + m_i + m_j \equiv \Delta_{fi} + m_i + m_j $
for different combinations of $ij$.
It is necessary to vacate the level $| n \rangle$ in order
to use the  population in the final state $| f \rangle$
lifted by laser as an experimental signature of the pair emission.
It is also highly desirable for unambiguous detection of the weak process,
$\gamma + | i \rangle \rightarrow \nu_i + \nu_j +
| f \rangle $ to measure a parity violating (PV) quantity such as
the rate difference due to different circular polarization of laser.

To explore all six neutrino pair thresholds from one metastable atom we need
to have an atom of the level structure of the corresponding complexity.
A strategy for determination of unknown neutrino parameters
shall be described elsewhere.
For our purpose of the relic neutrino detection,
observation of the pair
emission including the lightest neutrino, $m_1 + m_i \,(i = 1\,, 2\,, 3)$ 
near their mass thresholds is most important. 
Thus, effect of the neutrino form factor as given by
Fourier transformed atomic wavefunction ovelap,
$\langle n | e^{i(\vec{p}_1 + \vec{p}_2)\cdot \vec{x}} | i \rangle $, 
is not important, since the most significant region for
the relic neutrino effect appears in the small momentum region, 
$|\vec{p}_i| \ll $ the inverse of atomic size.
In the present work we shall also ignore PV effect, anticipating
that the PV quantity is comparable to parity-conserving quantity.

The rate via resonance is given by \cite{pair emission}
\begin{eqnarray}
&&
\Gamma^M(\omega\,; T_{\nu}) =
\frac{4G_F^2 F_0}{\pi \omega^2 \Delta \omega}
\frac{\gamma_r}{\gamma} \sum_{ij}
\theta (\omega - \Delta_{fi} - m_i - m_j ) \times
\nonumber \\ &&
\hspace*{1cm}
\int_{m_i}^{\omega - \Delta_{fi} - m_j} dE_1 I(E_1)
\left(1 - f_i(E_1) \right)
\left(1 - f_j(\omega - \Delta_{fi} - E_1) \right)
\,,
\\ &&
I(E_1) = k_0^{ij}
E_1 (\omega - \Delta_{fi} - E_1 )
\sqrt{(E_1^2 - m_i^2)\{ (\omega - \Delta_{fi} - E_1)^2 - m_j^2\}}
\nonumber \\ &&
+ k_M^{ij} \delta_{ij} m_i m_j 
\sqrt{(E_1^2 - m_i^2)\{ (\omega - \Delta_{fi} - E_1)^2 - m_j^2\}}
\,.
\label{rate formula}
\end{eqnarray}
Here $F_0$ is the photon number flux of laser light of
frequency resolution $\Delta \omega$.
Effect of the relic neutrino appears 
in the Pauli blocking factor of $(1 - f_i)(1 - f_j)$,
where $f_i $ is the Fermi-Dirac (FD) momentum distribution function
for the neutrino $\nu_i$.
We assumed the vanishing chemical potential in this formula;
the case of a finite chemical potential is discussed later.

The precise form of the distribution function after decoupling
follows time evolution equation in the expanding universe
with the Hubble rate $H(t) = \dot{a}(t)/a(t)$;
\begin{eqnarray}
&&
\left(
\frac{\partial}{\partial t} - \frac{\dot{a}(t)}{a(t)}p \frac{\partial}{\partial p}
\right)\,f(p\,; t) = 0
\,,
\end{eqnarray}
which has the solution of the form,
$f(p \,; t) = f(p a(t)/a(t_d)) = f(\,p(z_d+1)/(z+1)\,)$,
with $t_d$ the time of decoupling.
From this one concludes that the FD distribution in the present epoch
$t= t_0$, when written as a function of energy, takes the form of
\begin{eqnarray}
&&
f_i(E) = \frac{1}{e^{\sqrt{E^2 - m_i^2 + (m_i/(z_d +1))^2 }/T_{\nu}} + 1}
\,,
\end{eqnarray}
with $z_d$ the redshift factor at the neutrino decoupling.
The constants $k_a^{ij}$ at each threshold are functions of the neutrino mass
matrix elements whose explicit forms are in \cite{pair emission}.
In the case of the Dirac neutrino pair emission the rate formula is
modified and given by deleting the term $\propto k_M^{ij}$ in the above
formula,
which is the interference term of identical Majorana fermions.

Assuming a commercially available laser, one can take
the laser power (denoted above by $\omega F_0$) 
of order $1 W$ and the laser frequency
resolution $\Delta \omega / \omega \sim 10^{-9}$.
The threshold rise at $m_i + m_j$ is
of order, 
\begin{eqnarray}
&&
\Gamma_{ij}^M(\omega \,; 0) \sim
\frac{G_F^2 F_0 \gamma_r }
{\omega^2 \Delta \omega \gamma} 
(m_i m_j)^{3/2} (\omega - \Delta_{fi} - m_i - m_j)^2
(k_0^{ij} + k_M^{ij})
\label{threshold rate}
\\ &&
\approx
2.4 \times 10^{-22} s^{-1} (k_0^{ij} + k_M^{ij})
\frac{P}{W mm^{-2}}(\frac{eV}{\omega})^3
\frac{10^{-9}\omega}{\Delta \omega}\frac{10^{9}\gamma_r}{\gamma}
\nonumber \\ &&
\times \frac{(m_i m_j)^{3/2} (\omega - \Delta_{fi} - m_i - m_j)^2}{(0.1 eV)^5}
\,,
\end{eqnarray}
disregarding the relic effect, and 
\begin{eqnarray}
&&
\Gamma_{ij}^M(\omega \,; 0) \sim
\frac{G_F^2 F_0 \gamma_r }
{30 \omega^2 \Delta \omega \gamma} 
(\omega - \Delta_{fi} )^5
(k_0^{ij} + k_M^{ij})
\\ &&
\hspace*{-1cm}
\approx
8.1 \times 10^{-19} s^{-1} (k_0^{ij} + k_M^{ij})
\frac{P}{W mm^{-2}}(\frac{eV}{\omega})^3
\frac{10^{-9}\omega}{\Delta \omega}\frac{10^{9}\gamma_r}{\gamma}
(\frac{\omega - \Delta_{fi}}{1 eV})^5
\,,
\end{eqnarray}
far away from the threshold.
Here $\gamma = \sqrt{\gamma_i^2 + \gamma_n^2}$ is 
the width associated with initial and intermediate atomic levels,
both assumed metastable, for instance $1/\gamma > 1 sec$,
while $\gamma_r$ is E1 width of the final level $| f \rangle$
of order $1 ns$.
The angle factor is for instance $k_0^{11} + k_M^{11} =
2 \cos^2 \theta_{12}\cos^2 \theta_{13} \sim 1.3 $ at $2m_1$ threshold.

Effects of the relic neutrino are maximal near the laser energy
threshold of $\omega = \Delta_{if} + m_i + m_j$.
At this threshold both momenta of two emitted neutrinos vanish, and
the Pauli blocking factor becomes
\begin{eqnarray}
&&
(1 - f_i)(1 - f_j) = \frac{1}{(1 + e^{- m_i/T_d})(1 + e^{- m_j/T_d})} 
\sim \frac{1}{4} + \frac{m_i + m_j}{8 T_d}
\,.
\end{eqnarray}
Here
\begin{eqnarray}
&&
\frac{m_i}{T_d} \approx 5 \times 10^{-10} \frac{m_i}{1 meV}\frac{2 MeV}{T_d}
\,, \nonumber
\end{eqnarray}
with $T_d$ the neutrino decoupling temperature of order $2 MeV$
\cite{weinberg-cosmology}.
The theoretically calculated ratio 
\begin{eqnarray}
&&
r(\omega\,; T_{\nu}) \equiv \frac{\Gamma^M(\omega\,; T_{\nu})}{\Gamma^M(\omega\,; 0)}
\,,
\label{relic ratio}
\end{eqnarray}
thus approaches $\approx
\frac{1}{4} + \frac{m_i + m_j}{8 T_d}$ 
at the $m_i+m_j$ threshold.
The threshold rate is eq.(\ref{threshold rate}) times this ratio.

\begin{figure}[htbp]
  \centerline{\includegraphics[scale=0.8]{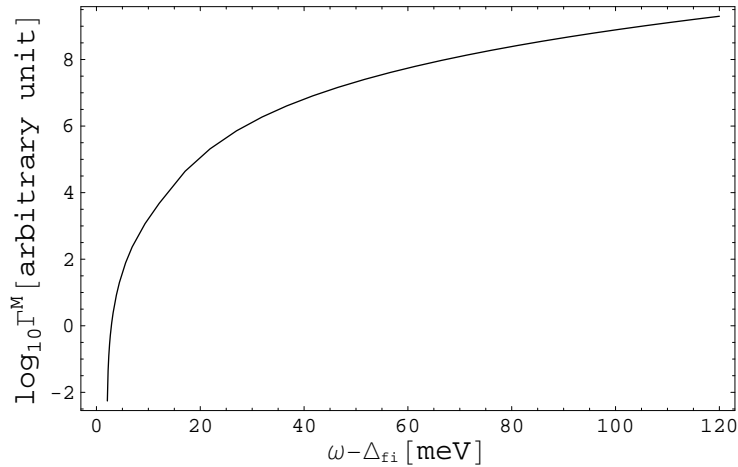}}
  \caption{Event rate with relic effect included}
  \label{Fig 2 }
\end{figure}

An experimental strategy is then as follows.
For a very small $m_i/T_d$ one derives as a function of 
$\omega - \Delta_{fi}$
the ratio of experimental data to
the theoretical value $\Gamma^M(\omega\,; 0)$, which is meant 
to be the theoretical rate without the Pauli blocking.
The theoretical function $\Gamma^M(\omega\,; 0)$ contains 
both mixing angle factors as an overall factor and the mass $m_i$.
One can determine both of these parameters internally 
from an experiment by fitting this
ratio normalized to $\approx 1/4$ at the threshold.
With an exteme precision one may even hope to measure the decoupling
temperature $T_d$.
We note that precision measurement of mixing angles is not a prerequisite
in this approach.

\begin{figure}[htbp]
  \centerline{\includegraphics[scale=0.8]{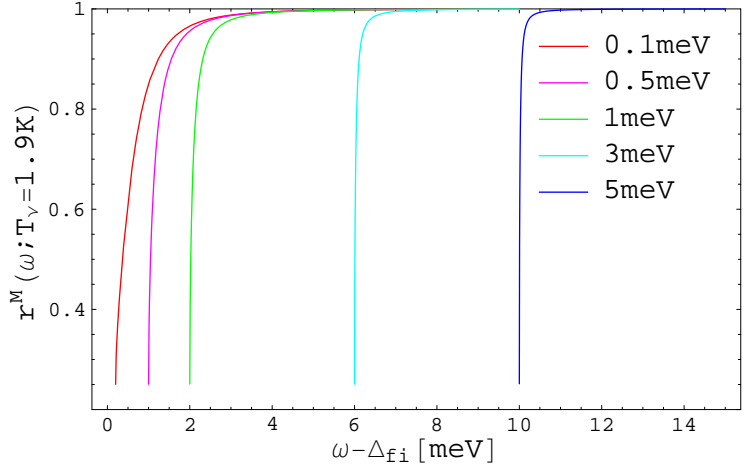}}
  \caption{Ratio with to without relic effect
  for several neutrino masses; $0.1 \sim 5 \,meV$}
  \label{Fig 3}
\end{figure}

In order to discuss the magnitude of relic neutrino effect,
we assume in the present work that all neutrino masses are known with a good
precision.
Numerical results for the rate $\Gamma^{M}(\omega\,; 1.9 K)$ 
in the laser energy range including $2m_1 \sim 2m_3$ 
are shown in Figure 2 for the Majorana case 
assuming the standard neutrino temperature $T_{\nu} = 1.9 K$.
We took for the neutrino parameters, $m_1 = 1meV$ and $\sin^2 \theta_{13}
= 0.032$, the maximal allowed value, and other paremeters
constrained by neutrino oscillation data.
The Dirac case can be dealt with in a similar way.

The distinction of Majorana and Dirac neutrinos is possible
at higher thresholds above $m_1 + m_2$, and shall be
explained elsewhere.
Here we shall assume that all neutrinos are of the Majorana type.

\begin{figure}[htbp]
  \centerline{\includegraphics[scale=0.9]{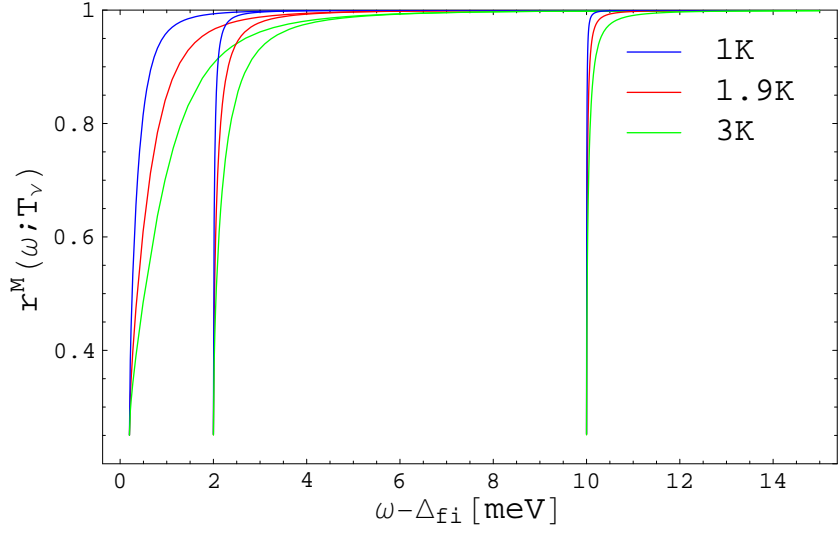}}
  \caption{Difference of relic effect for temperature variation
  taking 3 neutrino masses; 0.1 ,\, 1,\, 5 $meV$}
  \label{Fig 4 }
\end{figure}

\begin{figure}[htbp]
  \centerline{\includegraphics[scale=0.9]{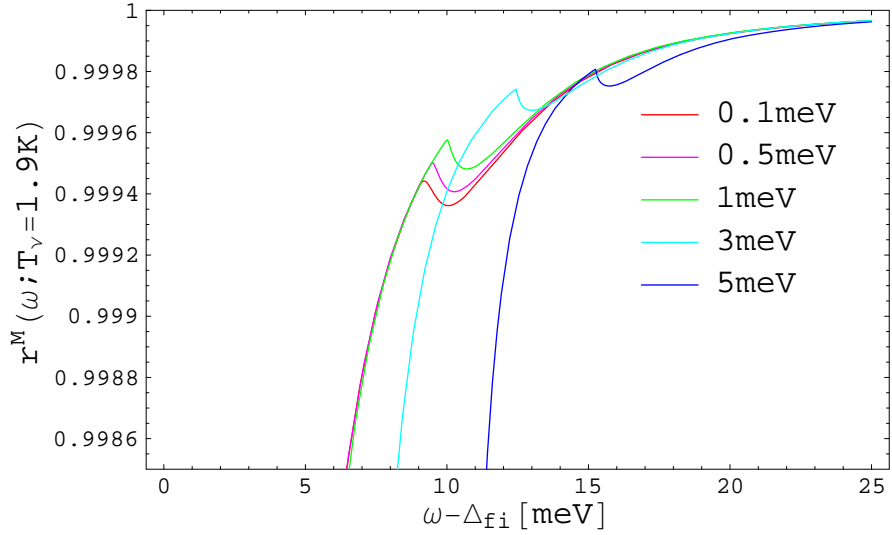}}
  \caption{Ratio with to without relic effect in $m_1 + m_2$ mass region
	for indicated $m_1$ values}
  \label{Fig 5 }
\end{figure}

The ratio $r(\omega\,; 1.9 K) $ (\ref{relic ratio}) 
near the $2m_1$ threshold region 
is shown in Figure 3 assuming the Majorana neutrino
of the mass range $m_1 = 0.1 meV \sim 5 meV$.
For smaller values of the neutrino mass $m_1$
a larger region in the laser energy exits for
visible relic neutrino effect.

In Figure 4 we plot the ratio $r(\omega\,; T_{\nu}) $ (\ref{relic ratio}) 
for different neutrino temperatures $T_{\nu} = (1 \sim 3) K$.
It appears that if high statistics data become available,
the temperature determination at the level of 10 \% is possible 
for smaller $m_1$ masses.

The region at higher thresholds is also interesting, because
the event rate is much larger, for instance by $O[10^3]$ at $m_1 + m_{2}$, 
than at $\omega - \Delta_{fi} = 3m_1$ near $2m_1$ threshold
($m_1 = 1meV$ taken).
This is shown in Figure 5 near $m_1 + m_2$ threshold, where
one needs a precision of $O[10^{-4}]$ for detection of the relic effect.

We shall now turn to implications of 
determination of the relic neutrino effect.
We discuss two issues; (1) limit on the chemical potential,
hence the lepton asymmetry,
(2) restriction on extra species of particles.

The difference between particles and anti-particles for the
Dirac neutrino, and two helicity states for the Majorana case, 
is reflected by a finite chemical potential $\mu$,
which is related to the lepton asymmetry.
For a small chemical potential the lepton asymmetry is
of order $\mu /T_{\nu}$;
\begin{eqnarray}
&&
\frac{n_L}{n_{\gamma}} \sim 
\frac{\pi^2}{12 \zeta(3)}\frac{\mu}{T_{\nu}} \approx 0.68 \frac{\mu}{T_{\nu}}
\,.
\end{eqnarray}
It is natural to expect the asymmetry of order, $\mu /T_{\nu} = O[10^{-10}]$,
the same order as the baryon asymmetry.
In leptogenesis scenario \cite{leptogenesis}
the lepton asymmetry $L$ of this order has a definite relation to
the baryon asymmetry $B$;
$L = -51B/28$, taking 3 generations and 1 Higgs doublet for
the standard model \cite{b-l relation}.
Hence observation of the lepton asymmetry is an unambiguous test of
leptogenesis scenario.
Although a measurement of $\mu/T_{\nu} = O[10^{-10}]$ effect is 
extremely difficult,
it would be  a rewarding challenge to verify or falsify the leptogenesis scenario.

\begin{figure}[htbp]
  \centerline{\includegraphics[scale=0.9]{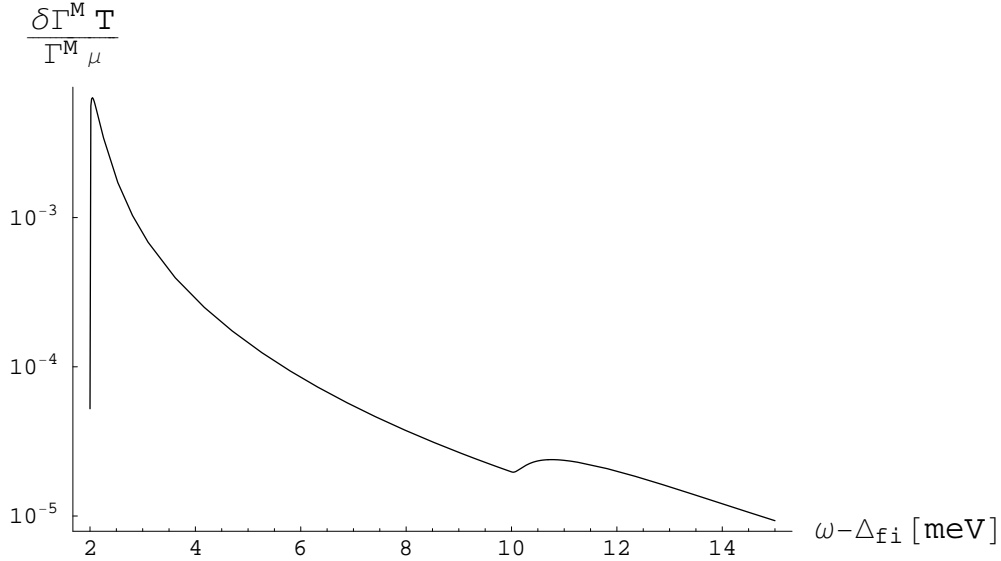}}
  \caption{Effect of finite chemical potential}
  \label{Fig 6 }
\end{figure}

From nucleosynthesis, one has a crude limit on the chemical
potential of all neutrino flavors $\alpha$ (considering the neutrino oscillation
is important in this respect \cite{nosci-cosmology}), of order
$|\mu_{\alpha}/T_{\nu}| \leq 0.04$, much larger than the expectation of
leptogenesis.
Thus, it would be interesting to improve the bound on 
the chemical potential $\mu$ from neutrino pair emission experiments.

The FD function in $z_d \rightarrow \infty$ limit takes the
form for different helicity $h$ states,
\begin{eqnarray}
&&
f(p\,; \mu) = \frac{1}{e^{(p + h\mu)/T_{\nu}} + 1}
\,.
\end{eqnarray}
To leading $\mu$ order, 
\begin{eqnarray}
&&
1 - f(p\,; \mu) \approx 1 -  f(p\,; 0) 
+ h \frac{\mu}{T_{\nu}}f(p\,; 0) 
\left(  1 -  f(p\,; 0) \right)
\,,
\end{eqnarray}
hence in neutrino helicity sums the linear term in helicity $h$
is relevant.
From corresponding formulas of \cite{correction} the leading linear term
in the chemical potential is thus derived as
\begin{eqnarray}
&&
\delta \Gamma^{M\,, D}(\omega\,; T_{\nu}) =  \frac{\mu}{T_{\nu}}
\frac{ G_F^2 F_0}{2\pi \omega^2 \Delta \omega}
\frac{\gamma_r}{\gamma} \sum_{ij}
\theta (\omega - \Delta_{fi} - m_i - m_j ) \times
\nonumber \\ &&
\hspace*{1cm}
\int_{m_i}^{\omega - \Delta_{fi} - m_j} dE_1 I_{\mu}(E_1)
(1 - f_i)(1 - f_j)
\,,
\\ &&
I_{\mu}(E_1) = k_0^{ij}
\sqrt{(E_1^2 - m_i^2)\{ (\omega - \Delta_{fi} - E_1)^2 - m_j^2\}} \times
\nonumber \\ &&
\hspace*{-1cm}
\left(
f_i (\omega - \Delta_{fi} - E_1 ) \sqrt{E_1^2 - m_i^2}
+ f_j E_1 \sqrt{ (\omega - \Delta_{fi} - E_1)^2 - m_j^2 }
\right)
\,,
\label{effect of chemical potential}
\end{eqnarray}
where the FD function $f_i$ refers to the form of zero
chemical potential.
The threshold rate of this quantity is calculated as
\begin{eqnarray}
&&
\delta \Gamma^{M\,, D}(\omega\,; T_{\nu}) \sim
\frac{\sqrt{2}}{30 \pi}
\frac{\mu}{T_{\nu}} \frac{ G_F^2 F_0}{ \omega^2 \Delta \omega}
\frac{\gamma_r}{\gamma} \times
\nonumber \\ &&
k_0^{ij} m_i m_j (\sqrt{m_i} + \sqrt{m_j})(\omega - \Delta_{fi} - m_i - m_j)^{5/2}
\,.
\end{eqnarray}
There is no difference between the Majorana and Dirac cases.
As an illustration, we show in Figure 6 effect of a finite chemical
potential plotting the quantity $\delta \Gamma^{M\,, D}(\omega\,; 1.9 K)T_{\nu}/
(\Gamma^{M}(\omega\,; 1.9 K) \mu)$
in the energy range $2m_1 \sim m_1 + m_2$.

The neutrino temperatue is also a sensitive probe for
physical processes after the neutrino decoupling.
We are content here to discuss a trivial implication once the neutrino
temperature is determined with a precision.
Suppose that hypothetical light $\Delta N_{eff}$ species of particles, 
either bosons or fermions, with weight factors 1 and 7/8
respectively,
exist, and are thermally coupled to $e^{\pm}$ and nucleons (or light nuclei)
in the cosmic temperature range between the neutrino decoupling
and some freeze-out temperature prior to $e^{\pm}$ pair annihilation.
The present neutrino temperature is then modfied from $(4/11)^{1/3} T_0$
to $(4/11)^{1/3} (1 + 2 \Delta N_{eff}/11)^{- 1/3} T_0$,
with $T_0 \approx 2.7 K$ the cosmic microwave temperature.
This way one may derive a constraint on the number of
extra light species $\Delta N_{eff}$.

\vspace{0.5cm}
In summary, we discussed a challenging proposal of
measuring the cosmic temperature of relic neutrino, and
mentioned how to experimentally test the leptogenesis scenario.


\begin{thebibliography}{99}
\bibitem{correction}
In relevant formulas in \cite{pair emission},
eqs.(16), (17),(19),(20), (21), helicities were 
inadvertently flipped;
correct formulas are obtained by replacing $h_2$ there by $-h_2$.
\end{thebibliography}
\end{document}